\documentclass[aps,prl,twocolumn,showpacs,superscriptaddress]{revtex4}

\usepackage{epsfig}
\usepackage{graphicx}
\usepackage{color}

\def\beq{\begin{equation}}
\def\eeq{\end{equation}}
\def\bea{\begin{eqnarray}}
\def\eea{\end{eqnarray}}
\def\nn{\nonumber}
\def\sss{\scriptscriptstyle}
\def\bd{B_d^0}

\def\ks{K_{\sss S}}
\def\kbar{{\overline{K^0}}}
\def\roughly#1{\mathrel{\raise.3ex\hbox
{$#1$\kern-.75em\lower1ex\hbox{$\sim$}}}}

\def\pew{P_{\sss EW}}
\def\pewc{P_{\sss EW}^{\sss C}}
\def\pewp{P'_{\sss EW}}
\def\pewcp{P_{\sss EW}^{\sss \prime C}}
\def\btopik{B \to \pi K}
\def\btopipi{B \to \pi\pi}
\def\btod{{\bar b} \to {\bar d}}
\def\btos{{\bar b} \to {\bar s}}
\def\btoq{{\bar b} \to {\bar q}}

\begin{document}

\preprint{UdeM-GPP-TH-06-149} 
\preprint{KEK-TH-1012}
\preprint{IMSc/2006/08/18}

\title{\boldmath Patterns of New Physics in $B$ Decays}

\author{Maxime Imbeault} 
\affiliation{Physique des Particules, Universit\'e de Montr\'eal, \\
C.P. 6128, succ.~centre-ville, Montr\'eal, QC, Canada H3C 3J7}
\author{David London} 
\affiliation{Physique des Particules, Universit\'e de Montr\'eal, \\
C.P. 6128, succ.~centre-ville, Montr\'eal, QC, Canada H3C 3J7}
\author{Chandradew Sharma} 
\affiliation{Institute of Mathematical Sciences, C. I. T Campus,
Taramani, Chennai 600 113, India}
\author{Nita Sinha} 
\affiliation{Institute of Mathematical Sciences, C. I. T Campus,
Taramani, Chennai 600 113, India}
\author{Rahul Sinha}
\affiliation{Institute of Mathematical Sciences, C. I. T Campus,
Taramani, Chennai 600 113, India} 
\affiliation{Theory Group, KEK, Oho 1-1 Tsukuba, 305-0801, Japan}

\date{\today}

\begin{abstract}
  We show that any new physics (NP) which affects $B$ decays with
  penguin contributions can be absorbed by redefinitions of the
  standard-model (SM) diagrammatic amplitudes. Hence, there are no
  clean signals of NP in such decays unless there is an accurate
  theoretical estimate of parameters or a justifiable approximation
  can be made. In all decays with penguin contributions, NP
  simultaneously affects pairs of diagrams. The evidence for a large
  $C'$ from fits to $\btopik$ data is naturally explained if NP
  contributes to $\pewp$, since NP affects the $\pewp$ and $C'$
  diagrams as a pair. The weak phase $\gamma$ measured in $\btopik$
  decays will always agree with its SM value even in the presence of
  NP, if the NP contributes in such a way that the amplitudes retain
  the SM form after suitable redefinitions.
\end{abstract}

\maketitle

It is convenient to use the diagrammatic description to compose the
amplitudes for $B$ decays. In this approach the amplitudes are written
in terms of eight diagrams: the color-favored and color-suppressed
tree amplitudes $T$ and $C$, the gluonic penguin amplitude $\cal P$,
the color-favored and color-suppressed electroweak penguin amplitudes
$\pew$ and $\pewc$, the annihilation and exchange amplitudes $A$ and
$E$, and the penguin-annihilation diagram $PA$ \cite{GHLR}. There are
also other diagrams, but they are much smaller and in general can be
neglected.

The penguin diagram ${\cal P}\!_q$ for $\btoq$ transitions (${\bar q}
= {\bar s}, {\bar d}$) receives contributions from each of the
internal quarks ${\bar u}$, ${\bar c}$ and ${\bar t}$ and hence does
not have a single well-defined weak phase.  However, using the
unitarity of the Cabibbo-Kobayashi-Maskawa (CKM) matrix, one can write
\begin{eqnarray}
{\cal P}\!_q & = & V_{ub}^* V_{uq} P_u + V_{cb}^* V_{cq} P_c
+ V_{tb}^* V_{tq} P_t \nn\\
& = & V_{ub}^* V_{uq} (P_u - P_c) + V_{tb}^* V_{tq} (P_t - P_c)~.  
\end{eqnarray}
The phase information in the CKM matrix is conventionally parametrized
in terms of the unitarity triangle, in which the interior CP-violating
angles are known as $\alpha$, $\beta$ and $\gamma$ \cite{pdg}. ${\cal
P}\!_q$ may be expressed in terms of these weak phases as
\begin{equation}
  \label{eq:b2q}
{\cal P}\!_s  \equiv  P'_{uc} e^{i\gamma} - P'_{tc} ~,~~
{\cal P}\!_d  \equiv  P_{uc} e^{i\gamma} + P_{tc} e^{-i\beta}~,
\end{equation}
where the negative sign of $V_{ts}$ has been explicitly factored out.
We use the convention that a prime on any diagrammatic amplitude
indicates a $\btos$ transition and distinguishes it from the $\btod$
transition.  $PA$, which contributes only to final states that are
self-conjugate at the quark level, is treated similarly. Note that,
technically, both $\pew$ and $\pewc$ also have two pieces. However,
the $u$-quark pieces for these two diagrams are proportional to the
$u$-quark mass, and are greatly suppressed. They will be neglected
here.  Hence all $B$-decay amplitudes can be written in terms of the
ten diagrams $T$, $C$, $P_{tc}$, $P_{uc}$, $\pew$, $\pewc$, $A$, $E$,
$PA_{tc}$ and $PA_{uc}$, each of which are associated with a
well-defined weak phase.

Symmetries such as isospin may be used to relate different $B$ decay
amplitudes and thus reduce the number of independent diagrammatic
amplitudes. Some of the decays where this approach is particularly
useful include $B\to\pi\pi$, $\btopik$, $B\to\rho\rho$, etc. In this
paper it is assumed that isospin is a good symmetry, even for any new
physics (NP) beyond the standard model (SM).

We next consider the contribution of NP to these decays. There are a
variety of ways of parametrizing this NP, but in the present paper we
propose a diagrammatic approach which allows each diagram to be
individually modified: $D e^{i\phi_{\sss SM}} \to D e^{i\phi_{\sss
SM}} + N\!P e^{i\phi_{\sss N\!P}}$. The addition of NP effects in
terms of diagrams is equivalent to the inclusion of NP operators in
terms of quarks~\cite{DLNP}. Now, NP matrix elements can in general
include several contributions, and may not have a well-defined weak
phase.  However, individual matrix elements do have well-defined
phases. We therefore consider one by one the addition of these
individual NP matrix elements to the diagrams.

Although our results are quite general, as an example we focus on
$B\to\pi K$ decays because present data are at odds with the
predictions of the SM \cite{Baek}. This discrepancy, often refrerred
to as the ``$K\pi$-puzzle,'' is not yet statistically significant as
it is only at the level of 2-3$\sigma$. It is nevertheless intriguing
since there are other disagreements in $\btos$ decays.  However, in
this paper we do present other $B$-decay amplitudes to illustrate our
points.

In the SM the four $\btopik$ amplitudes are given by
\begin{eqnarray}
A(B^+ \to \pi^+K^0) & = & -P' + P'_{uc} e^{i\gamma} ~,
\label{newbkpiamps}\\ 
\sqrt{2} A(B^+ \to \pi^0K^+) & = & (P' + \pewp + \pewcp) \nn\\
& & ~~~~-(T'+C'+P'_{uc}) e^{i\gamma}~, \nn\\
A(B^0\to \pi^-K^+) & = & (P' + \pewcp) - (T' +P'_{uc})
e^{i\gamma} ~, \nn\\
\sqrt{2} A(B^0\to \pi^0K^0) & = & -(P'- \pewp ) -(C'-P'_{uc})
e^{i\gamma} ~, \nn 
\end{eqnarray}
where, in order to simplify the expressions, we have introduced $P'
\equiv P'_{tc} -\frac13 \pewcp$ and absorbed $A'$ into other diagrams
by the redefinitions $P'_{uc}\to P'_{uc}+A'$, $T'\to T'-A'$, $C'\to
C'+A'$.

We now consider the addition of NP to $\btopik$ decays. For example,
consider first a NP contribution to the electroweak penguin: $\pewp
\to \pewp + N' e^{i\phi}$. Note that the Lorentz structure of the NP
piece is arbitrary. It has been shown that any complex number can be
written in terms of two other pieces with arbitrary phases
[reparametrization invariance (RI)] \cite{RI}. Since the $\btopik$
amplitudes involve the phases 0 and $\gamma$, we rewrite $ N'
e^{i\phi}$ as $N'_1 + N'_2 e^{i\gamma}$. With the addition of this NP
the amplitudes become
\begin{eqnarray}
\label{kpinp}
A(B^+ \to \pi^+K^0) & = & -P' + P'_{uc} e^{i\gamma} ~, \\
\sqrt{2} A(B^+ \to \pi^0K^+) & = & (P' + \pewp + \pewcp + N'_1) \nn\\
& & \!\!\!\!-(T'+C'-N'_2+P'_{uc}) e^{i\gamma}~, \nn\\
A(B^0\to \pi^-K^+) & = & (P' + \pewcp) - (T' +P'_{uc})
e^{i\gamma} ~, \nn\\
\sqrt{2} A(B^0\to \pi^0K^0) & = & -(P'- \pewp -N'_1 ) \nn\\
& & ~~~~-(C'-N'_2-P'_{uc}) e^{i\gamma} ~. \nn  
\end{eqnarray}
With the NP contributing to $\pewp$, only the amplitudes $A(B^+ \to
\pi^0K^+)$ and $A(B^0\to \pi^0K^0)$ are affected.  We can now remove
$N'_1$ and $N'_2$ by making the following redefinitions: $C' \to
\hat{C}' \equiv C' - N'_2$ and $\pewp \to \hat{P}'_{\sss EW} \equiv
\pewp + N'_1$. In this case, the above amplitudes now reduce to the
{\em same form} as in the SM [Eq.~(\ref{newbkpiamps})].  This has two
effects. First, we see that the NP cannot be detected directly through
measurements of $\btopik$ decays. Second, its only effect is to change
the size of $\pewp$ {\it and} $C'$. If the NP were to contribute to
$C' e^{i\gamma}$ instead of $\pewp$, the amplitudes in the presence of
this NP would again have the same form as in the SM with the
redefinitions $C' \to \hat{C}' \equiv C' + N'_2$ and $\pewp \to
\hat{P}'_{\sss EW} \equiv \pewp - N'_1$.

If the NP is added to other diagrams, the effect is similar. In all
cases the new amplitudes can be cast in the same form as that of the
SM. Hence there can be {\it no clean signal of NP} in the $\btopik$
modes. There are two cases. (i) The NP is added to $T'$:
$T'e^{i\gamma} \to T'e^{i\gamma} + N' e^{i\phi}$. $A(B^+ \to
\pi^0K^+)$ and $A(B^0\to \pi^-K^+)$ are affected. One makes the
redefinitions $T' \to \hat{T}' \equiv T' + N'_2$ and $\pewcp \to
\hat{P}^{\sss \prime C}_{\sss EW} \equiv \pewcp - N'_1$. The situation
is the same, apart from a change of relative sign in the
redefinitions, if the NP is added to $\pewcp$. (ii) The NP is added to
$P'$: $P' \to P' + N' e^{i\phi}$. Now all four amplitudes are
affected. One makes the redefinitions $P' \to \hat{P}' \equiv P' +
N'_1$ and $P'_{uc} \to \hat{P}'_{uc} \equiv P'_{uc} - N'_2$. Once
again the situation is the same, apart from a different relative sign
in the redefinitions, if the NP is added to $P'_{uc}$.

We therefore conclude that, regardless of how one adds the NP, it is
always possible to cast the $\btopik$ amplitudes as in the SM. This
indicates that there is technically no clean signal of NP in $\btopik$
decays. The fundamental reason for this is reparametrization
invariance \cite{RI} and the fact that the $\btopik$ amplitudes
involve two phases.

Similar results can be seen in $B \to \pi\pi$ decays. Using
the (re)definitions: $P\equiv P_{tc} -\frac13\pewc+PA_{tc}$ and
$P_{uc}\to P_{uc}+E+PA_{uc}$, the complete SM amplitudes for these
decays are
\begin{eqnarray}
- \!\sqrt{2} A(B^+\! \rightarrow \pi^+ \pi^0) &\!\!=\!\!& \left( T\! +\! C
\right) e^{i \gamma}\! +\left( \pew \!+ \pewc \right) e^{- i \beta}  ,~ \nn\\
- A(B^0\! \rightarrow \pi^+ \pi^-) &\!\!=\!\!& (T\! +\! P_{uc} ) e^{i
  \gamma}  \! +(\pewc\! +P) e^{- i \beta} ,\\
- \!\sqrt{2} A(B^0\! \rightarrow \pi^0 \pi^0) &\!\!=\!\!& (C\! -\! P_{uc} )
  e^{i \gamma} \!   +(\pew\! -P) e^{-
  i \beta} .\nn
\label{pipi}   
\end{eqnarray}
It is straightforward to show that if NP enters any of the diagrams
above, it can be divided into two pieces as $N e^{i\phi} = N_1
e^{i\gamma} + N_2 e^{-i\beta}$, and absorbed through redefinitions of
the SM diagrams. Hence no clean NP signal is possible in the $B \to
\pi\pi$ either. Once again, just as in the $\btopik$ case, we find
that NP introduced in $\pew$ or $C$ modifies $\pew$ and $C$ amplitudes
simultaneously.  The case for the pairs $(T,\pew)$ and $(P,P_{uc})$
are similar.

This said, it must be acknowledged that there are, in fact,
significant differences between $\btopipi$ and $\btopik$
decays. Within the SM, the $\pew$ and $\pewc$ contributions in
$\btopipi$ decays are expected to be tiny and are hence justifiably
ignored. If there is NP, the amplitudes cannot be recast in the SM
form if $\pew$ and $\pewc$ are neglected. In this case, direct CP
violation in $B^+ \rightarrow \pi^+ \pi^0$ is a clean signal of
NP. Thus, it is only if $\pew$ and $\pewc$ are kept that one can
conclude that no clean NP signal is possible.

Other $B$ decays can be analyzed similarly. The easiest way is to look
at the quark-level topologies. $\btopik$ and $\btopipi$ decays are
${\bar b} \to {\bar s} u {\bar u}$ and ${\bar b} \to {\bar d} u {\bar
u}$, respectively. The analysis of all decays of the form ${\bar b}
\to {\bar d} u {\bar u}$, ${\bar b} \to {\bar s} u {\bar u}$, ${\bar
b} \to {\bar d} c {\bar c}$, or ${\bar b} \to {\bar s} c {\bar c}$ is
thus identical to those above. If all SM diagrams are kept, any NP
effects can be absorbed by redefining the SM diagrams, so that there
are no clean signals of NP in such decays.

Decays of the form ${\bar b} \to {\bar d} d {\bar d}$, ${\bar b} \to
{\bar s} d {\bar d}$, ${\bar b} \to {\bar d} s {\bar s}$, or ${\bar b}
\to {\bar s} s {\bar s}$ are slightly different because $T$ and $C$
cannot enter (i.e.\ they are ``pure penguin''). For example, consider
$B^0\to K^0 \kbar$. The amplitude for this decay is
\begin{equation}
A(B^0 \to K^0 \kbar)  =  P_{uc}  e^{i\gamma}+~P 
e^{-i\beta} ~,
\end{equation}
where $P$ and $P_{uc}$ are as defined above except there is no $E$
contribution here.  It is easy to see that NP added to any of the
diagrams can once again be absorbed by redefinitions of the
amplitudes.  Thus, there can be no clean signal of NP in $B^0\to K^0
\kbar$ decays.

We therefore see that there are {\em no clean signals of NP in $B$
decays which receive penguin contributions}. This is the first point
of this paper.  This result was earlier stated in Ref.~\cite{LSS1999}.
However, NP was introduced there only by modifying one of the
contributing weak phases.  Here the conclusion is more robust as it is
obtained when an arbitrary NP amplitude is introduced. The only
possible way to have a signal of NP in modes with penguin
contributions is to make an approximation in the SM amplitudes. The
approximation must be such that NP-altered amplitudes cannot be recast
in the SM form. {\em The signal of NP so obtained will be reliable, if
the approximation is justifiable}.

In comparison, clean signals of NP are possible in the pure tree-level
decays ${\bar b} \to {\bar u} c {\bar d}$, ${\bar b} \to {\bar c} u
{\bar d}$, ${\bar b} \to {\bar u} c {\bar s}$, and ${\bar b} \to {\bar
c} u {\bar s}$.  These decays have only one weak phase and so NP can
be detected through measurements of direct CP violation.  However, any
NP effects are purely tree-level and are therefore much suppressed
\cite{DLN}.

In modes with penguin contributions, our observation that NP-modified
amplitudes retain the SM form has two additional consequences which we
discuss in detail.

It has been speculated that a large electroweak-penguin amplitude
resulting from NP could resolve the $K\pi$ puzzle. Several fits to the
$\btopik$ data \cite{Baek} seem to indicate that the
electroweak-penguin amplitude is indeed large, but they also find that
it is difficult to accomodate the data without demanding a
larger-than-expected $C'$ amplitude. A priori one would not expect a
tree-level amplitude to be substantially affected by NP. However, as
we have illustrated, the simultaneous effect of NP on $C'$ and $\pewp$
is not an accident. In fact, a second conclusion that emerges from the
above discussion is that {\em NP always affects diagrammatic
amplitudes in pairs}. To be specific, we find that the following pairs
of (suitably-redefined) amplitudes are simultaneously affected by NP:
\begin{itemize}
\item $\pewp$ and $C'$ ($\pew$ and $C$),
\item $\pewcp$ and $T'$ ($\pewc$ and $T$),
\item $P'$ and $P'_{uc}$ ($P$ and $P_{uc}$).
\end{itemize}
These pairs of diagrams are topologically equivalent in the sense that
if they contribute to any decay mode, they always appear together.
Hence it should have been expected that NP must contribute to specific
pairs of amplitudes simultaneously. The fact that fits to the
$\btopik$ data require a large $C'$ may well indicate a contribution
of NP to $\pewp$.

Now, the $\btopik$ modes have been used to measure the weak phase
$\gamma$.  It is important to examine the effect of NP on this
weak-phase measurement. We have shown that all the isospin-related
$\btopik$ amplitudes have the same form as in the SM even in the
presence of NP.  Hence, in principle, the weak phase measured using
these modes with or without NP should remain the same.  However, a
simple counting of the number of theoretical parameters versus the
number of independent observables implies that $\gamma$ cannot be
extracted without resorting to some approximations. As noted above,
when no approximations are made, NP does not alter the SM form of the
amplitudes. However, if some of the diagrammatic amplitudes are
neglected, then the NP-modified amplitudes may not have the SM form.
Only when the approximations made are such that the amplitudes with
the addition of NP retain the SM form, then the {\em weak phase
measured remains unaltered from its SM value, even in the presence of
NP}. This is our third result and it explains why fits to the
$\btopik$ data \cite{Baek} often yield $\gamma$ consisitent with the
CKM fits \cite{CKMfits}. In case the approximations made are such that
the amplitudes in the presence of NP cannot be recast in the SM form,
the addition of NP will result in adding to the number of theoretical
parameters, rendering the weak-phase measurement impossible without
further assumptions. If $\gamma$ is nevertheless measured and is found
to differ from the SM value, the deviation may be either due to NP or
due to the invalid approximations used. In this sense a discrepency in
the measured value of $\gamma$ is not necessarily an unambiguious
signal of NP.

The $\btopik$ modes are described by a larger number of parameters
than the possible measurements. There can at best be 9 independent
observables: the branching ratios and direct CP asymmetries for each
of the four decay modes and one time-dependent CP asymmetry for
$\bd\to \pi^0K^0$. Using Eq.~(\ref{newbkpiamps}), we see that there
are six diagrammatic amplitudes and two weak phases $\gamma$ and
$\beta$, resulting in a total of 13 parameters. The phase $\beta$ can
be taken from the measurement of $\sin 2\beta$ in $B^0(t) \to
J/\psi\ks$: $\sin 2\beta = 0.726 \pm 0.037$ \cite{sin2beta}, leaving
12 theoretical unknowns. It is hence clear that one needs to make some
approximation in order to analyze these modes.

In Ref.~\cite{GHLR}, the relative sizes of the amplitudes were
estimated to be roughly
\bea
1 : |P'_{tc}| ~~,~~~~ {\cal O}({\bar\lambda}) : |T'|,~|\pewp|
~,~~~~~~~~ \nn\\
{\cal O}({\bar\lambda}^2) : |C'|,~|P'_{uc}|,~|\pewcp| ~~,~~~~ {\cal
O}({\bar\lambda}^3) : |A'|,
\label{hierarchy}
\eea
where ${\bar\lambda} \sim 0.2$. These SM estimates are often used as a
guide to neglect diagrammatic amplitudes and reduce the number of
parameters.

If one retains only the ``large'' diagrams ($P'_{tc}$, $T'$, $\pewp$),
the $\btopik$ amplitudes can be written
\bea
A(B^+ \to \pi^+K^0) & \simeq & -P'_{tc} ~, \nn\\
\sqrt{2} A(B^+ \to \pi^0K^+) & \simeq & P'_{tc} + \pewp -T'
e^{i\gamma}~, \nn\\
A(B^0\to \pi^-K^+) & \simeq & P'_{tc} - T' e^{i\gamma} ~, \nn\\
\sqrt{2} A(B^0\to \pi^0K^0) & \simeq & -P'_{tc} + \pewp ~.
\label{smnoc}
\eea
Using the above approximate form, signals of NP are possible since
some of the amplitudes have a well-defined phase and RI is lost. For
example, the direct CP asymmetry in $B^+ \to \pi^+K^0$ and $B^0\to
\pi^0K^0$ vanishes. Similarly, the mixing-induced CP asymmetry in
$B^0(t) \to \pi^0K^0$ should be equal to that in $B^0(t) \to
J/\psi\ks$ (modulo a sign). Any deviation from these expectations may
imply presence of NP. Note, however, that this conclusion would be
based on the validity of the approximations made in the above
parametrization.

There are other examples where NP signals can be obtained under
approximations. The sum rule \cite{sumrule}
\bea
&& \Gamma(B^0\to \pi^-K^+) + \Gamma(B^+ \to \pi^+K^0) \approx \nn\\
&&  \hskip10truemm 2\Gamma(B^+ \to \pi^0K^+)+2\Gamma(B^0\to \pi^0K^0) ~.
\eea
holds approximately in the SM with errors that are quadratic in the
subdominant terms ($T'$, $\pewp$). The breakdown of this sum rule is a
signal of NP. 

Another approximate parametrization for the $\btopik$ amplitudes
considered in literature \cite{koy} is
\begin{eqnarray}
  \label{eq:kpi-2}
  A(B^+ \to \pi^+K^0) & \simeq & -P'_{tc} ~, \nn\\
\sqrt{2} A(B^+ \to \pi^0K^+) & \simeq & P'_{tc} + \pewp -(T'+C')
e^{i\gamma}~, \nn\\
A(B^0\to \pi^-K^+) & \simeq & P'_{tc} - T' e^{i\gamma} ~, \nn\\
\sqrt{2} A(B^0\to \pi^0K^0) & \simeq & -P'_{tc} + \pewp-C' e^{i\gamma} ~.
\end{eqnarray}
$C'$ is retained even though it is subdominant, as it is claimed that
the fit is extremely poor without retaining $C'$. Leaving aside the
merits of such an assumption, we note that in Eq.~(\ref{eq:kpi-2}) we
have 9 variables and 9 observables. The system of equations. can thus
be solved to obtain $\gamma$ and other variables. It is easy to see
that if NP contributes to $\pewp$ or $C'$ it can be reabsorbed using
RI. This is because $\pewp$ and $C'$ appear simultaneously in the
amplitudes. Since the amplitudes have the same form with or without NP
the value of $\gamma$ measured under this approximation {\em would not
differ from the SM value even in the presence of NP}.  If a
disagreement in the value of $\gamma$ is found, it must reflect a
failure of the assumptions made. In case NP contributes to other
topologies it {\em cannot} be reabsorbed using RI and the value of
$\gamma$ cannot be measured unless further approximations are made.

Given that the pairing of $\pewp$ and $C'$ is the reason for fits to
the $\btopik$ data finding a value for $\gamma$ consistent with the SM
CKM fits, it is puzzling why fits to the $\btopik$ data using
Eq.~(\ref{smnoc}) (e.g., Ref.~\cite{baeketal}) also result in a value
for $\gamma$ in agreement with the SM. The explanation lies in the
large error in the data, due to which the two parametrizations in
Eqs.~(\ref{smnoc}) and (\ref{eq:kpi-2}) cannot be distinguished. This
argument is vindicated by the fits performed in Ref.~\cite{baeketal}
when NP is included. If NP is added to $\pewp$ in Eq.~(\ref{smnoc}),
it can be recast in the form of Eq.~(\ref{eq:kpi-2}) with $C'$ being
replaced entirely by a NP contribution and $\pewp$ being
redefined. Only NP of this kind (scenario (i) of Ref.~\cite{baeketal})
results in the best fits and $\gamma$ is remarkably consistent with
the SM. The point is that the agreement of $\gamma$ obtained with the
SM value does not a priori rule out NP, but rather strengthens the
arguments in favor of NP contributing to $\pewp$.

To summarize, we have made three main points. First, there are no
clean signals of new physics (NP) in any $B$ decay which receives
penguin contributions. In order to obtain a signal of NP, it is
necessary to either have an  accurate theoretical estimate of
parameters or to make a justifiable approximation. Second, we have
noted that in all decays with penguin contributions, NP always affects
diagrammatic amplitudes in pairs. The diagrams $\pewp$ and $C'$,
$\pewcp$ and $T'$, and $P'$ and $P'_{uc}$ (with or without primes) are
simultaneously affected by NP. Fits to $\btopik$ data suggest
larger-than-expected $\pewc$ and $C'$ contributions. In view of our
observation, the requirement of large $C'$ may be a sign of
NP. Finally, we have shown that if NP contributes in such a way that
the amplitudes retain the SM form using RI, the weak phase obtained
will not be altered due to the presence of NP.  This provides a
natural explanation of the result that several fits to $\btopik$ data
with varying approximations yield $\gamma$ in accord with SM.  The
observation of a large $C'$ and $\gamma$ consistent with the SM in
$\btopik$ decays provide substantial circumstantial evidence in favor
of NP.

\vskip2truemm D.L. thanks F.~J.~Botella and J.~P.~Silva for helpful
conversations about $\btopik$ decays and reparametrization invariance.
D.L. also acknowledges a useful email exchange with M. Gronau and J.
Rosner. The work of M.I. and D.L. was financially supported by NSERC
of Canada. N.S. thanks the theory group, KEK for their hospitality,
where part of this work was done. The work of N.S. was partially
supported by the Department of Science and Technology , India.

\end{document}